\documentclass[a4paper,11pt,twocolumn]{article}

\usepackage[utf8]{inputenc}
\usepackage[T1]{fontenc}
\usepackage{amsmath}
\usepackage{amssymb}
\usepackage{graphicx}
\usepackage{braket}
\usepackage{color}
\usepackage[superscript,biblabel,nocompress,nospace]{cite}
\usepackage{units}
\usepackage{lmodern}
\usepackage{scrextend}

\usepackage{setspace}
\usepackage[pdftex,bookmarks,colorlinks,breaklinks]{hyperref}  
\hypersetup{linkcolor=red,citecolor=blue,filecolor=dullmagenta,urlcolor=blue} %

\usepackage[switch, modulo]{lineno}

\graphicspath{{./pics/}}

\changefontsizes{9pt}

\begin{document}
\twocolumn[
\huge\textbf{\textsf{Highly efficient frequency conversion with bandwidth compression of quantum light}}\vspace{0.5cm}

\large\textbf{\textsf{Markus Allgaier$^{1,*}$, Vahid Ansari$^1$, Linda Sansoni$^1$, Viktor Quiring$^1$, Raimund Ricken$^1$, Georg Harder$^1$, Benjamin Brecht$^{1,2}$, and Christine Silberhorn$^1$}}\vspace{0.5cm}
]

\footnotetext[1]{Integrated Quantum Optics, Applied Physics, University of Paderborn, 33098 Paderborn, Germany}
\footnotetext[2]{Clarendon Laboratory, Department of Physics, University of Oxford, Oxford OX1 3PU, United Kingdom}
\renewcommand{\thefootnote}{\fnsymbol{footnote}}
\footnotetext[1]{e-mail: markus.allgaier@uni-paderborn.de}

\textbf{\textsf{Hybrid quantum networks rely on efficient interfacing of dissimilar quantum nodes, since elements based on parametric down-conversion sources, quantum dots, color centres or atoms are fundamentally different in their frequencies and bandwidths. While pulse manipulation has been demonstrated in very different systems,
to date no interface exists that provides both an efficient bandwidth compression 
and a substantial frequency translation at the same time. Here, we demonstrate an engineered sum-frequency-conversion process in Lithium Niobate that achieves both goals. We convert pure photons at telecom wavelengths to the visible range while compressing the bandwidth by a factor of 7.47 under preservation of non-classical photon-number statistics. We achieve internal conversion efficiencies of 75.5\%, significantly outperforming spectral filtering for bandwidth compression.
Our system thus makes the connection between previously incompatible quantum systems as a step towards usable quantum networks.}} 

Photons play the important role of transmitting quantum information between nodes in a quantum network \cite{kimble_quantum_2008}. However, systems employed for different tasks such as generation, storage and manipulation of quantum states are in general spectrally incompatible. 
Therefore, interfaces to adapt the central frequency and bandwidth of the photons are crucial \cite{lavoie_spectral_2013, fisher_frequency_2016, karpinski_bandwidth_2016}. To achieve any viable bandwidth-compression the interface has to provide at least a net gain over using spectral filters. 
Electro-optical frequency conversion can provide such high efficiencies for bandwidth compression \cite{karpinski_bandwidth_2016} and shearing \cite{wright_spectral_2016} of quantum pulses. However, it is limited to frequency shifts of a few hundred GHz. 
Optical frequency conversion in nonlinear crystals offers both large frequency shifts as well as high conversion efficiencies \cite{vandevender_high_2004, albota_efficient_2004, pelc_long-wavelength-pumped_2011, bre2014a, baune_strongly_2015}.  
Operating on chirped pulses allows to perform spectral shaping \cite{raymer_manipulating_2012}, an approach with which a bandwidth compression of 40 has been demonstrated \cite{lavoie_spectral_2013}, however, with low efficiencies below spectral filtering.
Reaching high conversion efficiencies with this method is challenging, since very broad phasematching is required, which in turn limits the allowed interaction lengths and hence the conversion efficiencies. 
An alternative approach is to engineer the phasematching of the sum-frequency process itself \cite{baronavski_analysis_1993} by choosing appropriate group velocity and pump-pulse conditions. Such engineering has been widely exploited for parametric down-conversion \cite{mosley_heralded_2008, jin_widely_2013, ben_dixon_spectral_2013, bruno_pulsed_2014} to produce decorrelated photon pairs efficiently. For frequency conversion, this approach has not been investigated. However, we show in this work that dispersion engineering can be used to develop processes that provide spectral reshaping and high conversion efficiencies at the same time.

The quantum pulse gate (QPG) \cite{chr2011,bre2014a,ans2016, manurkar_multidimensional_2016} is such a device that exploits specific group-velocity conditions: The input and the pump are group-velocity matched, while the output is strongly group-velocity mismatched. 
This is achieved in a type-II sum-frequency process in a periodically poled Titanium-indiffused waveguide in Lithium Niobate.
The group velocity matching ensures that spectrally broad input pulses overlap throughout the crystal while the mismatch with the output in combination with the long interaction length inside the waveguide results in a narrow output spectrum. 
Furthermore, the output temporal mode, i.e. the temporal or spectral amplitude of the output pulse, only depends on the phasematching and not on the pump or input fields. This allows to convert any input to the same narrow output. It can thus interface broad PDC sources as well as narrower and even dissimilar emitters such as quantum dots.

To demonstrate the performance of the QPG as an interface, we focus on its application as a link between PDC sources and quantum memories to produce on-demand single photons.
Ideally for quantum networks, single photons are generated into well defined optical modes and feature compatibility with low-loss fibre networks.
Heralded photons from engineered, single-pass PDC fulfill these requirements \cite{eckstein_highly_2011,har2013}. 
One class of quantum memories, Raman quantum memories, can exhibit very broad spectral bandwidths of a few GHz \cite{sprague_broadband_2014} up to 20\,GHz \cite{poem_broadband_2015, hepp_electronic_2014}. Long storage times have been achieved in alkali vapour memories with bandwidths of up to several GHz \cite{saunders_cavity-enhanced_2016}, however these are narrowband compared to the above mentioned PDC sources with bandwidths in the THz regime \cite{har2013}.
In diamond, THz bandwidth can be achieved \cite{england_photons_2013}, but both storage time and memory efficiency are low, such that these memories cannot be utilised in quantum networks, yet.
In principle, schemes exist to match both systems directly by using a very broad memory \cite{england_photons_2013} or strong spectral filtering of a correlated PDC source \cite{michelberger_interfacing_2015}, but these come at the expense of short storage times or reduced purities through spectral filtering \cite{christ_probing_2011}. 
An bandwidth-compressing interface between the broadband PDC sources at telecom wavelengths and the narrower quantum memories at visible or near infrared wavelengths is therefore very desirable. 
We demonstrate such an interface by converting single photons from 1545nm and a bandwidth of 1\,THz to 550\,nm and a bandwidth of 129\,GHz under preservation of the second order correlation function \(g^{(2)}(0)\) while achieving external conversion efficiencies high enough to outperform a spectral filter producing an equivalent output spectrum.

\begin{figure}
\centering
\includegraphics[width=0.5\textwidth]{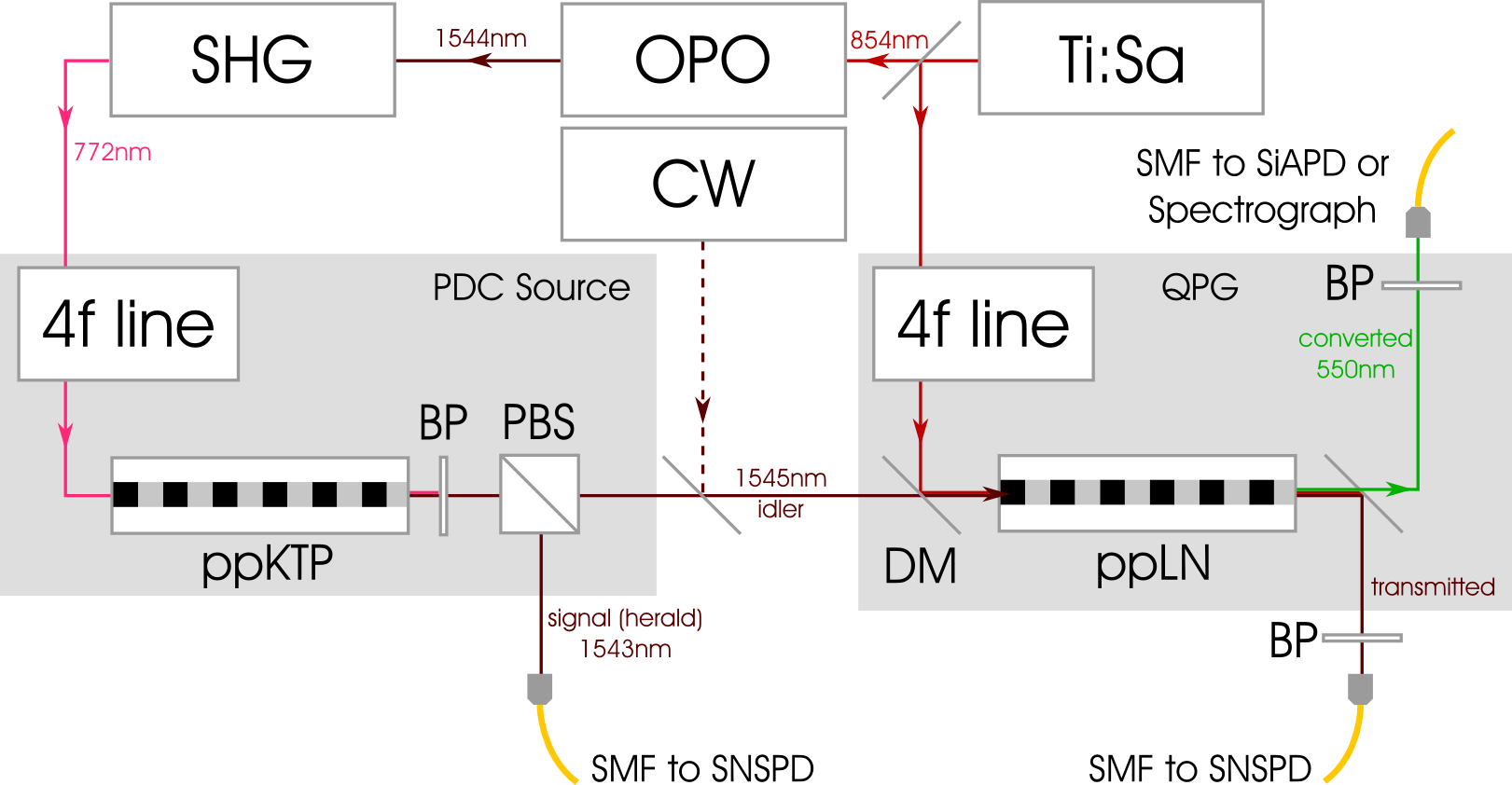}
\caption{Setup used for characterization of the QPG's transfer function as well as the measurement of conversion efficiency and spectra. BP: Band pass filter, PBS: Polarizing beam splitter, DM: Dichroic mirror, SMF: Single mode fibre, MMF: Multi mode fibre, SNSPD: Superconducting nanowire single photon detector, SiAPD: Silicon avalanche photo diode, CW: Tunable continuous wave telecom laser}
\label{fig:setup}
\end{figure}

\begin{figure}
\centering
\includegraphics[width=0.22\textwidth]{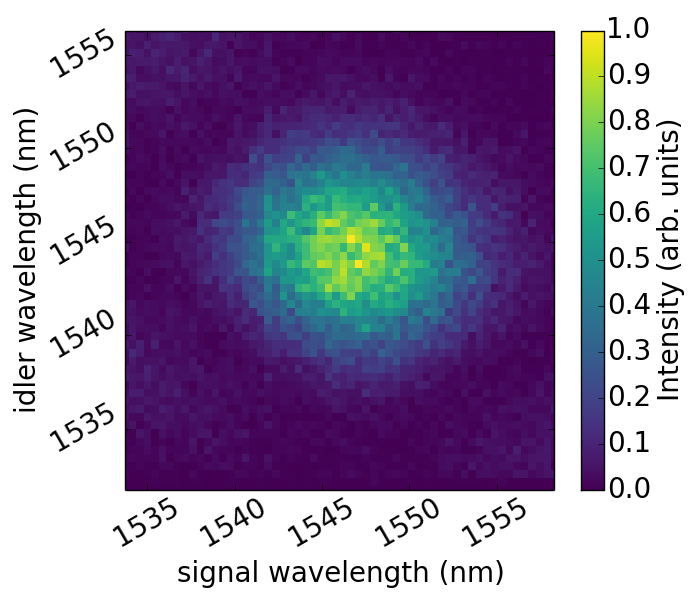}\includegraphics[width=0.25\textwidth]{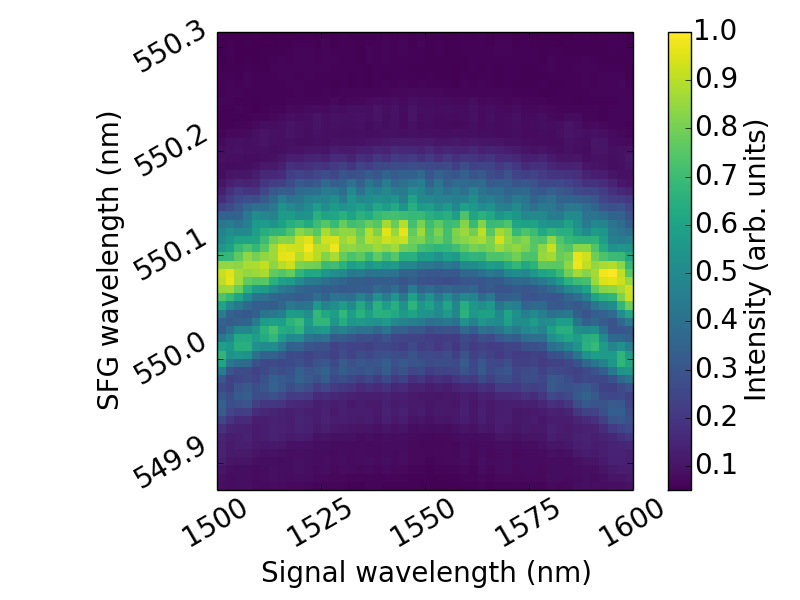}
\caption{Left: Joint spectral intensity of the PDC photons. Right: Phasematching function of the QPG.}
\label{fig:jsi}
\end{figure}

Our experimental setup is depicted in Fig. \ref{fig:setup}. 
We generate single photons at 1545\,nm from a type-II parametric down-conversion source (PDC) in periodically poled Potassium Titanyl Phosphate (KTP).
The pump beam for the PDC source is created by a series of elements, starting with a Ti:sapphire mode-locked laser, which pumps an optical parametric oscillator (OPO), followed by second harmonic generation (SHG) and bandwidth fine-tuning with a 4-f spectral filter. The bandwidth is adjusted to ensure a decorrelated PDC state. 
We characterize the spectral properties of the PDC photons by measuring their joint spectral intensity (JSI) with a time-of-flight spectrometer \cite{avenhaus_fiber-assisted_2009}, consisting of a pair of dispersive fibres and a low-jitter superconducting nanowire single-photon detector (Photon Spot SNSPD). From this measurement, shown in figure \ref{fig:jsi} (left), we conclude that the bandwidth (FWHM) of the signal photon 963\(\pm\)11\,GHz at 1545\,nm central wavelength. Furthermore, the round shape of the JSI indicates that the photon pairs are indeed spectrally decorrelated.
We keep the pulse energy of the PDC pump at a low level of 62.5\,pJ to ensure that mainly photon pairs and only few higher-photon-number components are created. At the output, an 80\,nm wide band-pass filter centred at 1550\,nm is used to filter out background processes while not cutting the spectrum of the actual PDC process. 

The heralded signal photon is then sent to the QPG, consisting of a 27\,mm long LiNbO\(_3\) crystal with Ti-indiffused waveguides. It is pumped at 854\,nm with light from the same Ti:Sapphire laser, which is spectrally shaped by means of another 4f-line containing a liquid-crystal spatial-light modulator (SLM). 
The SLM can be used to adapt the QPG pump to any input. 
To characterize the QPG, we measure its phasematching function by recording the sum-frequency signal of a broad pump and a tunable continuous-wave (CW) telecom laser on a Czerny-Turner spectrograph equipped with 2398 lines/mm grating and a single-photon-sensitive emCCD camera. The result is shown in the right plot of Fig. \ref{fig:jsi}. The horizontal orientation of the phasematching function is due to the fact that the input and pump are group-velocity matched, while the output is strongly group-velocity mismatched. This leads to the narrow spectrum of the output field while accepting a broad input field. As the curvature of the phasematching function is connected to the group-velocity mismatch between input and pump, the horizontal portion on the top indicates perfect group velocity matching, where the output spectrum depends only on the phasematching and not on the pump. This holds for a telecom input bandwidth as large as 20\,nm. Since the PDC photons are only 7.8\,nm wide, we are well within that range and adjust the pump bandwidth accordingly to ensure maximum conversion efficiency. After the conversion, we separate both the converted and the unconverted light from the background and residual pump by means of broadband filters and couple all fields into single-mode fibres. It is noteworthy that the phasematching bandwidth and therefore the bandwidth compression depend on the sample length and could therefore be increased or decreased to get the desired output. As the group-velocity curves steepen towards shorter wavelength, moving the process in this direction would increase the group-velocity mismatch between input and output resulting in even greater bandwidth compression.

To be viable as an interface in quantum networks the device has to leave the quantum nature of the single photons untouched. To measure this we employ photon-number statistics, namely the heralded second-order autocorrelation function of signal photons measured with a 50/50 beam splitter and two click detectors:
\begin{equation}
g^{(2)} = \frac{P_{cc}}{P_{1}\cdot P_{2}},
\end{equation}
where $P_{cc}$ is the coincidence probability and $P_{1}$ and $P_{2}$ the single-click probabilities.
The QPG does not change the \(g^{(2)}(0)\), which takes the value of 0.32\(\pm\)0.01 both before and after the frequency conversion. With \(g^{(2)}(0)<1\) and constant, the single-photon character is verified before and after the conversion.

\begin{figure}
\centering
\includegraphics[width=0.45\textwidth]{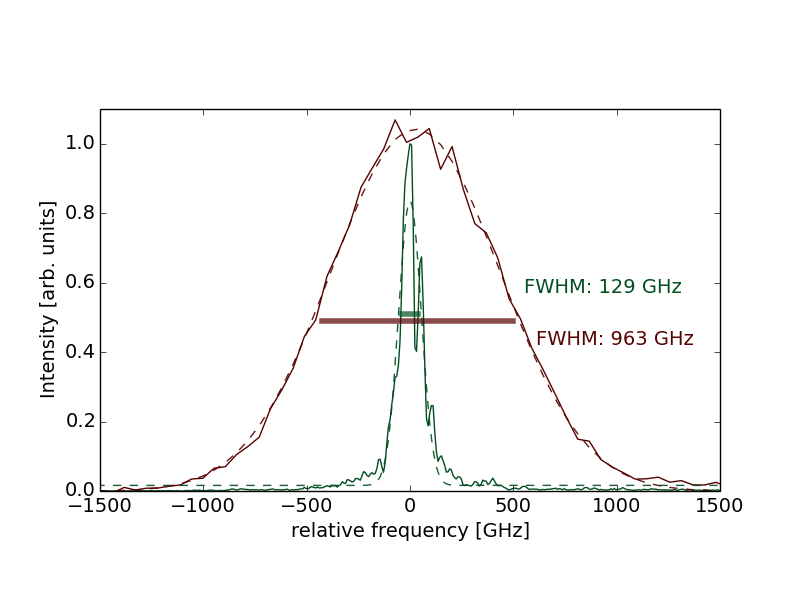}
\caption{Marginal spectra of the idler PDC photon before (red) and after (green) frequency conversion centred around their respective centre frequencies. Dashed lines correspond to gaussian fits.}
\label{fig:spectra}
\end{figure}

To estimate the bandwidth compression, we record the spectrum of converted PDC photons with the aforementioned Czerny-Turner spectrometer. The marginal spectra of the signal photon together with the converted spectrum are depicted in figure \ref{fig:spectra}. The converted light has a spectral bandwidth of 129\(\pm\)4\,GHz and a central wavelength of 550\,nm. Compared to the original bandwidth of 963\,GHz of the PDC photon this implies a bandwidth-compression factor of 7.47\(\pm\)0.01. 

The second, equally important figure of merit is the conversion efficiency. If the conversion efficiency is low, a simple spectral filter could outperform the device. Were the signal converted by a CW pump, the bandwidth would remain constant at 963\,GHz. Filtering down to 129\,GHz would then imply a throughput of 13.40\(\pm\)0.02\,\% (the error corresponds to the uncertainties for the spectral bandwidths measured here), assuming the conversion itself is lossless. 
To measure the conversion efficiency, we send the photons to SNSPDs or and a silicon avalange photodiode (SiAPD) for infrared or visible photons, respectively. 
We estimate the internal efficiency of the process itself as well as the external efficiency including all optical loss in the setup.
As a measure for the internal efficiency we use the depletion of the signal by calculating the Klyshko-efficiency \cite{klyshko_use_1980} $\eta_\mathrm{t}$ of the unconverted 1545\,nm light, transmitted through the QPG with the QPG pump open and blocked. The Klishko-Efficiency is defined as $\eta_t = P_\mathrm{cc}/P_\mathrm{h}$, where $P_\mathrm{cc}$ is the coincidence-count probability between the herald and unconverted PDC photon and $P_\mathrm{h}$ is the herald-count probability alone. 
From this depletion we get the internal conversion efficiency of the process
\begin{equation}
\eta_\mathrm{int} = 1 - \frac{\eta_\mathrm{t}^\mathrm{open}}{\eta_\mathrm{t}^\mathrm{blocked}}.
\end{equation}
Using the depletion of the unconverted light has the advantage that it provides a direct measure of the internal conversion efficiency. By contrast one would need precise knowledge of all losses to estimate it from the up-converted signal. The resulting value for the internal conversion efficiency is 75.5\,\%. 

As a measure for the external conversion efficiency we use the ratio between the Klyshko-efficiencies of the converted light \(\eta_c\) and the unconverted light before the QPG \(\eta_u\), corrected only for the different detection efficiencies of the SiAPD compared to the SNSPD:
\begin{equation}
\eta_\mathrm{ext} = \frac{\eta_c\cdot\eta_\mathrm{SNSPD}}{\eta_u\cdot\eta_\mathrm{SiAPD}}
\end{equation}
This external conversion efficiency is 16.9\,\%.
Due to some spatial mode mismatch the coupling of the converted light into a single mode fibre is reduced compared to the unconverted light. Taking into account this reduced fibre compatibility of the green mode (50\,\% instead of 80\,\% for the herald) the external efficiency amounts to 27.1\,\%. This can be seen as the free-space efficiency of the device. As all of these efficiencies result from counting sufficiently large numbers of photons, errors are negligible. The coupling of the green mode into the fibre can be further improved  by optimizing the waveguide structure or the coupling optics. The difference between the internal and the external efficiency is mainly due to linear-optical losses in uncoated lenses and a 4-f-line bandpass filter with a total transmission of 68\,\% and a waveguide-incoupling efficiency of around 71\,\%.

These conversion efficiencies show that the QPG offers useful bandwidth compression and provides a net gain over using a spectral filter. For the first time this is realised in combination with substantial frequency conversion. This is true not only when looking at the internal conversion efficiency but even when comparing to the external conversion efficiency, which already includes all losses, like waveguide and even fibre couplings.

\begin{figure}
\centering
\includegraphics[width=0.25\textwidth]{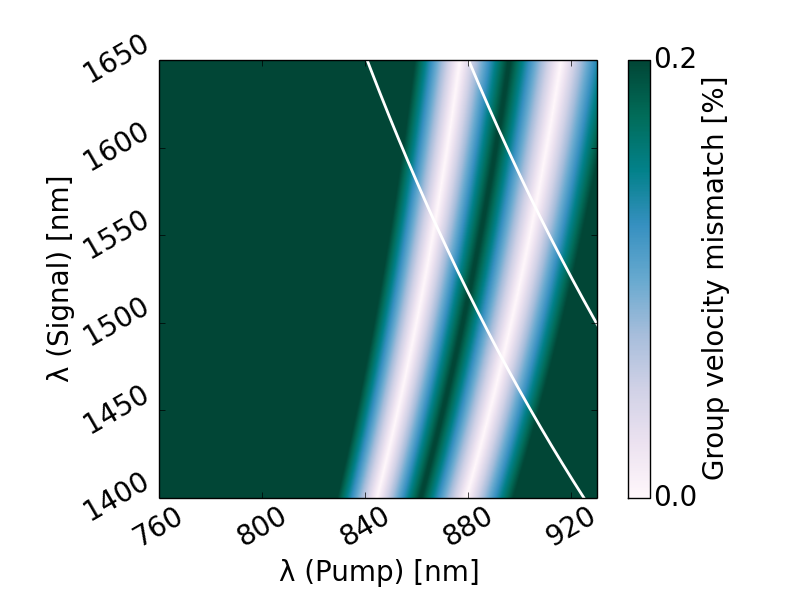}\includegraphics[width=0.25\textwidth]{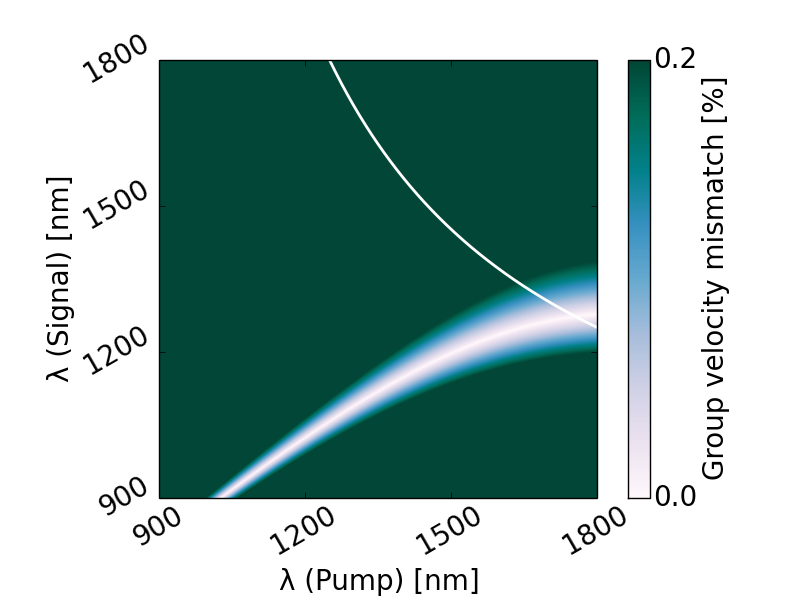}
\caption{Left: Group velocity matching in Ti:LiNbO3 between waveguide modes in ordinary and extraordinary polarization at two different temperatures (left stripe: 190\(^\circ\)\,C, right stripe: 300\(^\circ\)\,C). The solid white lines indicate wavelength combinations where the SFG process reaches the desired wavelength of 574\,nm (right line) for the NV0-transition in diamond and the wavelength of 550\,nm (left line) chosen in this letter. Right: Group velocity matching in bulk LiTaO3 the ordinary (o) and extraordinary (e) polarization at 190\(^{\circ}\)C. Here, the white line indicates an output wavelength of 738\,nm, corresponding to the SV-transition in diamond.}
\label{fig:lnprocess}
\end{figure}

Having demonstrated a viable interface, we calculate the process parameters required to interface the proposed broadband memories in diamond based on nitrogen and silicon vacancy centres \cite{poem_broadband_2015, hepp_electronic_2014}. The degrees of freedom available for tuning the conversion process are primarily the temperature and the choice of the nonlinear material.  
As a basis for this study we use effective Sellmeier equations \cite{edwards_temperature-dependent,jundt_temperature-dependent_1997} of the modes inside the waveguide. Figure \ref{fig:lnprocess} (left) shows the group-velocity mismatch between signal and pump at two different temperatures. The two light stripes in the color code represent areas with zero group velocity mismatch for 190\(^{\circ}\)C (left stripe) and 300\(^{\circ}\)C (right stripe), whereas the solid white lines indicate wavelength combination where the sum-frequency is at the desired output frequency. 
The main target wavelength in this work, the NV0 transition \cite{poem_broadband_2015} at 574\,nm, can be addressed with a group velocity matched SFG process at a sample temperature of 300\(^{\circ}\)C.
The signal wavelength would be at 1560\,nm and the pump at 907\,nm, well within reach of PDC sources and Ti-sapphire laser systems. For the proof-of-principle experiment in this work we have chosen a slightly different operating point of 190\(^{\circ}\)C as it simplifies the choice of suitable ovens and insulation materials, thus shifting the target wavelength to 550\,nm. As unwanted effects like photorefraction are only present at lower temperatures\cite{augstov_temperature_????}, there is no fundamental limitation for increasing the temperature as high as Curie temperature. 
The alternative silicon-vacancy transition \cite{hepp_electronic_2014} at 738\,nm cannot be reached with the birefringent properties of Lithium Niobate. However, Lithium Tantalate, a less birefringent material, supports it. 
The signal wavelength of this process could be at 1278\,nm and the pump wavelength at 1748\,nm, or vise versa. Temperature tuning of the group velocity matching in the same way as in Lithium Niobate can also be considered. Figure \ref{fig:lnprocess} (right) shows the parameter space for that process. Note that these numbers are based on bulk Sellmeier equations \cite{abedin_temperaturedependent_1996} and might slightly differ for waveguides.
Apart from sum-frequency processes, difference-frequency generation can also be considered. For example, conversion of near-infrared light as emitted by semiconductor quantum dots to the telecom band can be done with an infrared pump such as the one employed in Ref. \cite{pelc_downconversion_2012}. Overall, a large range of wavelengths can be covered with the available materials and realistic process parameters.

In conclusion we have realised a device that not only offers efficient upconversion from telecom light to the visible spectrum but also useful bandwidth compression. As the phasematching bandwidth is proportional to the inverse of the sample length, the compression factor is in principle scalable. It is noteworthy that the device does not provide a fixed bandwidth ratio between input and output but rather a fixed output bandwidth, such that the same converter can be used for inputs of different bandwidth.

This work was funded by the Deutsche Forschungsgemeinschaft via SFB TRR 142 and via the
Gottfried Wilhelm Leibniz-Preis.

MA and VA carried out the experiment. MA wrote the manuscript with support from GH. VQ and RR fabricated the LiNbO3 sample. LS, GH and BB helped supervise the project. BB and CS conceived the original idea, CS supervised the project.


\bibliography{own,telecom,memories,bandwidth,theory,materials,Georg}
\bibliographystyle{naturemag}

\end{document}